\documentclass[twocolumn,aps,prl,showpacs, superscriptaddress, amsmath,amssymb]{revtex4}

\usepackage{graphicx}
\usepackage{dcolumn}
\usepackage{bm}

\begin{document}

\title{Experimental study of the jamming transition at zero temperature}
\author{Xiang Cheng}
\altaffiliation{Current address: Department of Physics, Cornell
University, Ithaca, New York 14853, USA} \email{xc92@cornell.edu}
\affiliation{The James Franck Institute and Department of Physics,
The University of Chicago, Chicago, Illinois 60637, USA}

\date{\today}

\begin{abstract}

We experimentally investigate jamming in a quasi-two-dimensional
granular system of automatically swelling particles and show that a
maximum in the height of the first peak of the pair correlation
function is a structural signature of the jamming transition at zero
temperature. The same signature is also found in the second peak of
the pair correlation function, but not in the third peak, reflecting
the underlying singularity of jamming transition. We also study the
development of clusters in this system.  A static length scale
extracted from the cluster structure reaches the size of the system when
the system approaches the jamming point.  Finally, we show that in a
highly inhomogeneous system, friction causes the system to jam in
series of steps. In this case, jamming may be obtained through
successive buckling of force chains.

\end{abstract}

\maketitle

\section{I. Introduction}

\begin{figure}
\begin{center}
\includegraphics[width=3.35in]{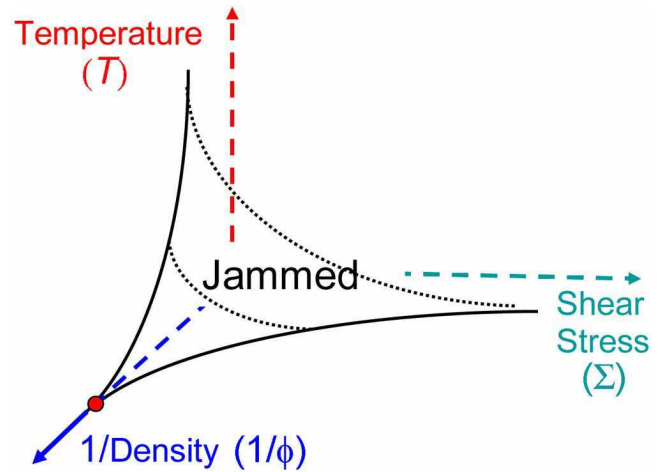}
\end{center}
\caption[Jamming phase diagram]{(Color online) Jamming phase
diagram \cite{Liu,O'Hern}. The volume bounded by the black lines
near the origin is the jammed phase. When a material crosses the
black line from the outside, it goes through the jamming transition.
The jamming point along the $1/\phi$ axis, point J, is marked by a
red dot.} \label{Figure1}
\end{figure}

Glasses return to the liquid state upon heating - they become soft
and can flow.  Sand flowing through a pipe or out of an orifice can
easily jam and become rigid.  In fact, large classes of materials,
ranging from polymer melts to foams, from glasses to dense colloidal
suspensions and granular matter, show a similar transition between a
flowing liquid-like state and a non-equilibrium, disordered solid
state.  How to understand the nature of this transition is one of
the central questions in different fields of materials science
\cite{Anderson,Ediger,Eisenberger,Weeks,Jaeger}.  The recently
proposed jamming phase diagram provides an approach to unify this
effort \cite{Liu,O'Hern}.  In such a phase diagram
(Fig.~\ref{Figure1}), athermal systems such as granular media sit in
the inverse-density($1/\phi$)/shear-stress($\Sigma$) plane, and
thermal systems, such as glass-forming liquids, sit in the
inverse-density($1/\phi$)/temperature($T$) plane.  When decreasing
the temperature, decreasing the shear stress or increasing the
density, a system goes from an unjammed phase into a jammed state.
At the jamming transition, a material becomes rigid and loses its
ability to explore the entire phase space efficiently; it therefore
falls out of equilibrium.

Due to its clean and clear signature, jamming along the $1/\phi$
axis in the jamming phase diagram, at zero temperature and zero
shear stress, has so far attracted the most attention in simulations
and theoretical studies
\cite{O'Hern,Silbert1,Silbert2,Ellenbroek,Somfai,Shundyak,Zeravcic,Henkes,Mailman,Wyart1,Wyart2}.
At the zero-temperature jamming point for frictionless spheres
(called point J) one observes the onset of rigidity along with other
interesting phenomena such as the divergence of the pair correlation
function and the appearance of anomalous soft modes
\cite{O'Hern,Silbert1,Silbert2,Ellenbroek,Somfai,Shundyak,Zeravcic,Mailman}.
It has been hypothesized that the properties of point J will
influence the jamming transition also nearby, similar to critical
points in second order phase transitions.  Indeed, it has been found
recently that the signature of the jamming transition at zero
temperature shows its vestige at finite temperature \cite{Zhang,Xu}.
Although there has been intensive theoretical and simulation work,
only a few experiments have been conducted to study the nature of
the zero-temperature jamming transition
\cite{Majmudar,Jacob,Clement}. Especially, the structural signature
of the jamming transition has not been directly addressed in
experiments by far. Corwin {\it et al.} probed the structural
signature of the jamming transition indirectly at zero temperature
along the $\Sigma$ axis of the jamming phase diagram \cite{Corwin}. Can
one directly observe the structural signature of a zero-temperature
jamming transition along the $1/\phi$ axis experimentally?  How is
the picture of jamming modified in a real system with frictional
interactions?  This paper investigates these essential questions.

Here we study the zero-temperature jamming transition experimentally
in a quasi-two-dimensional granular system of macroscopic particles.
By continuously and uniformly increasing the packing fraction, the
system is forced to go through the zero-temperature jamming point.
In this process, we find a structural signature, which is shown as a
maximum of the height of the first peaks of pair correlation
function.  A similar signature was previously seen in a colloidal
sample at non-zero effective temperature \cite{Zhang,Xu}.  Here, we
observe a maximum in the second peak as well and, by measuring the
pressure on the boundary of system, we show that these features
coincide with the onset of rigidity.  We also find that {\it en
route} to the jammed phase, our athermal system always
self-organizes itself into a structure consisting of particle
clusters.  A static length scale can be extracted from this
structure; it varies between a few particles to the size of the
system when the jamming point is approached.  Finally, we show that
friction can result in multiple jamming points in the presence of
highly heterogeneous particle arrangements. A system with friction
shows a clear historical dependence. Frictional effects are shown to
be reduced or eliminated when small amplitude vibrations are
introduced.


\section{II. Experimental method}

\begin{figure}
\begin{center}
\includegraphics[width=3in]{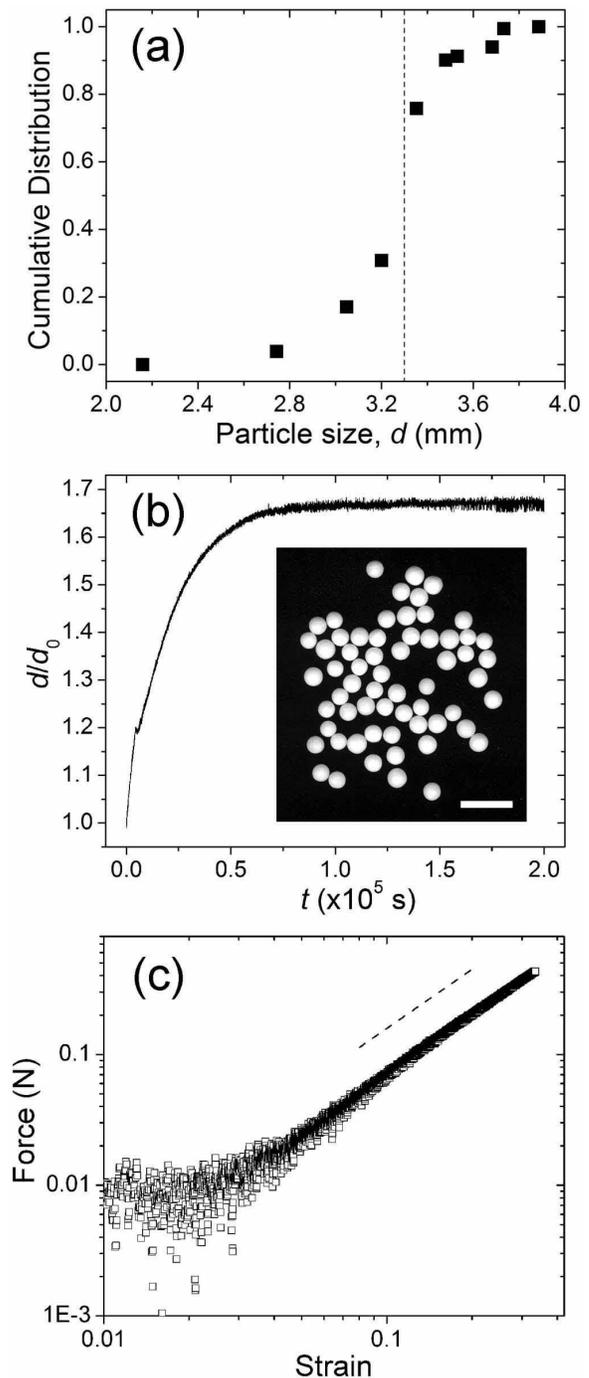}
\end{center}
\caption[Properties of tapioca pearls]{Properties of tapioca pearls.
(a) Cumulative distribution of diameter of dry tapioca pearls. The
vertical dashed line marks the average diameter of pearls. (b)
Diameter of a swelling pearl under water relative to its original
size as a function of time. Inset: optical image of dry tapioca
pearls. The white scale bar is 1 cm. (c) Compression force on a
fully-swelled tapioca pearl as a function of its strain. The strain
is defined as $\Delta x/d$, where $\Delta x$ is the deformation of
the particle under compression and $d=4.7$ mm is the original size
of the swelled particle at zero compression. The slope of the dashed
line is 3/2.} \label{Figure2}
\end{figure}

The granular material used consists of tapioca pearls, which are
spherical particles made of starch (Fig.~\ref{Figure2}b inset).  The
average diameter of {\it dry} pearls is 3.3mm and the cumulative
distribution of the sizes of dry pearls is shown in
Fig.~\ref{Figure2}a.  The polydispersity of the pearls is $8\%$.
Such a spread in size is essential to avoid crystallization in a two-dimensional (2D)
system.  One important property of tapioca pearls is that, when
submerged in water, they uniformly expand in size
(Fig.~\ref{Figure2}b).  The final diameter of a tapioca pearl can be
1.7 times larger than its original value. The swelling process is
very slow.  It takes 24 hours for a pearl to reach its final, fully
swelled state. Thus, the system is quasi-static.  During swelling,
particles keep their approximately spherical shape. The contact
interaction between {\it fully} swelled particles is purely
repulsive and of Hertzian type (Fig.~\ref{Figure2}c). The strain
versus stress curve on an individual particle is measured with an
Instron System (Model 5869): a particle is put between two
horizontal metal plates and the normal force on the top plate is
measured while the gap between the two plates decreases. For
particles which are not fully swelled, the inter-particle contact is
still purely repulsive but may deviate from Hertzian form. Some
particles may also show slightly plastic deformation.

\begin{figure}
\begin{center}
\includegraphics[width=3.35in]{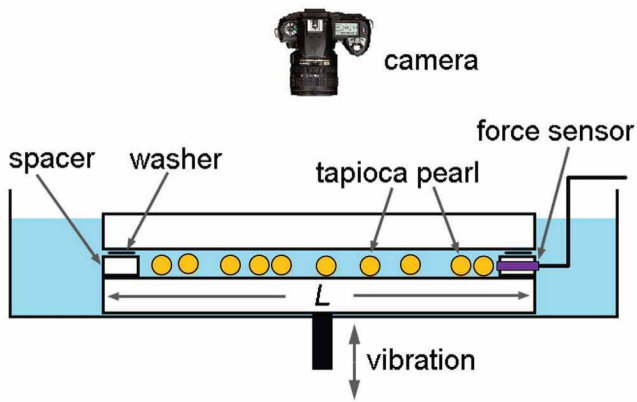}
\end{center}
\caption[Schematics of the experimental setup]{(Color online)
Schematics of the experimental setup. Shown here is the
cross section of the cell. The blue (gray) area indicates water.}
\label{Figure3}
\end{figure}

At the beginning of an experimental run, we put tapioca pearls
randomly into a square cell submerged in water. The schematics of
the setup can be seen in Fig.~\ref{Figure3}.  Two different sizes of
cells are used.  The side length of the larger cell is $L=54.6$ cm,
which can hold over 15,000 particles with initial packing fractions
$\phi_\mathrm{initial} \gtrsim 0.63$ (among which about 10,000
particles are studied in the central area to avoid the boundary
effect); the side length of the smaller cell, which can contain
roughly 2,000 particles with about 1,000 particles in the central
area, is $L=16.5$ cm. In the small cell, we installed a force sensor
(Futek load cell, Model LSB200) along one of its sides so that the
force or pressure along that boundary can be measured. For both the
large and small cells, the gap between the top and bottom plates is
kept by a 5.2 mm spacer enclosing the entire boundary.  To allow
water to flow in and out of the cell, a few thin washers (0.254 mm
in thickness) are put on top of the spacer.  The total gap thickness
(spacer + washer = 5.454 mm) prevents fully swelled particles from
buckling out of the plane significantly to form two layers. The cell
can also be coupled to a mechanical shaker at the bottom, which can
vibrate the entire cell vertically. However, unless explicitly
indicated, the experiments described here are done without shaking.
Vibration is only introduced at the end of Sec. V for controlling
the friction between particles.  An image of the particles is
recorded by a camera mounted above the cell.  For a system to reach
its fully jammed stationary state, a typical experiment takes
between 17 and 24 hours.  An image was taken every 20 s so that 3000
to 4000 images were recorded for each experiment. From these images,
the center and trajectory of individual particles were extracted.

We track the center of particles based on an algorithm developed for
colloidal systems \cite{Crocker}.  The images obtained are first
processed with a band-pass filter to eliminate any global intensity
variation and the pixel-size noise.  This is a standard procedure
before tracking the center of particles \cite{Crocker}.  The
resulting images are then convoluted with a mask consisting of a
white disk with radius a little smaller than the size of particles.
This eliminates any intensity variation on the top surface of
individual particles. The local maxima are then located.  The center
of particles can be found more precisely by calculating the centroid
of a blob around each local maximum \cite{Crocker}.  The center of
particles found in this way is more accurate at early times or low
packing fractions.  When the system is deep inside jammed phase, the
interface between particles has much lower contrast.  Therefore, the
error of particle tracking becomes larger.  When $\phi>0.90$, about
$1-2\%$ of particles are missed by the algorithm.

There are several advantages of this system.  First, different from
other granular systems for studying the jamming transition, where
the packing fraction is changed by either changing the number of
particles or by changing the volume of system from the boundary
\cite{Majmudar,Dauchot,Lechenault,Keys}, here we can continuously
and uniformly increase the packing fraction across the entire
system. Second, the system is quasi-static due to the extremely slow
swelling of the particles.  This allows the static structure of the
pack to be easily investigated.  Third, after swelling tapioca
pearls are much softer than other commonly used granular materials
such as glass beads.  By assuming the Poisson's ratio of particles
around 1/3, we can estimate the Young's modulus of swelled
particles from the force-strain curve shown in Fig.~\ref{Figure2}c.
The Young's modulus of swelled tapioca pearls is $0.060\pm0.005$
GPa, which is three orders of magnitude smaller than that of glass
beads. Thus the system can reach far inside the jammed phase.  This
is essential to directly see any small structural signatures of the
jamming transition. With hard granular materials, structural
signatures can only be probed indirectly by measuring the contact
force distribution between particles \cite{Corwin}. Also, since the
entire system is under water, the friction between particles and the
bottom plate of the cell is reduced due to lubrication. It is
interesting to note that our system is a 2D version of the old
experiment done by Stephen Hales in 1727 \cite{Hales,Zallen}.  In
order to find out how many contacting neighbors a spherical particle
has in a dense pack, Hales put peas into a fixed volume container
full of water and counted how many dimples each pea had after they
had swelled.


\section{III. Structural signature of jamming transition at $T=0$}

\begin{figure}
\begin{center}
\includegraphics[width=3.35in]{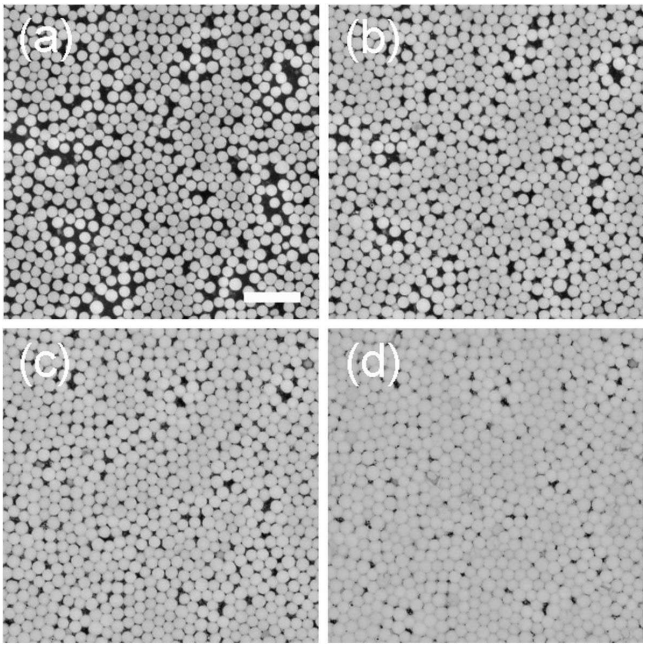}
\end{center}
\caption[Jamming transition at zero temperature]{Jamming transition
at zero temperature. The diameter of the particles, and therefore
the packing fraction, increases with time. (a) $\phi=0.62$ at $t=0$
h, (b) $\phi=0.76$ at $t=1.10$ h, (c) $\phi=0.84$ at $t=2.81$ h, and (d)
$\phi=0.92$ at $t=16.66$ h. The white scale bar is 2 cm.}
\label{Figure4}
\end{figure}

We initially prepared the 2D samples in a dense, but unjammed phase
(Fig.~\ref{Figure4}a). As the particles become larger, the packing
fraction, $\phi$, of the pack increases uniformly across the entire
system (Fig.~\ref{Figure4}b and \ref{Figure4}c). At a certain moment the system
crosses the jamming point and goes into the jammed phase
(Fig.~\ref{Figure4}d).  The jamming transition appears to be
continuous.  The question posed here is whether one can identify the
jamming point by looking merely at the structure of the pack.

\subsection{A. Experimental results}

\begin{figure}
\begin{center}
\includegraphics[width=3.35in]{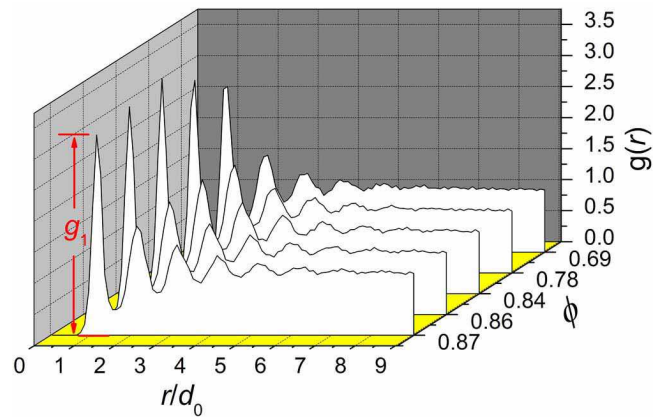}
\end{center}
\caption[Pair correlation, $g(r)$, at different packing
fractions]{(Color online) Pair correlation, $g(r)$, at different
packing fractions. The distance $r$ is given in the unit of the
average dry particle diameter, $d_0$. The height of the first peak
of $g(r)$, $g_1$, is indicated.} \label{Figure5}
\end{figure}

\begin{figure}
\begin{center}
\includegraphics[width=3.35in]{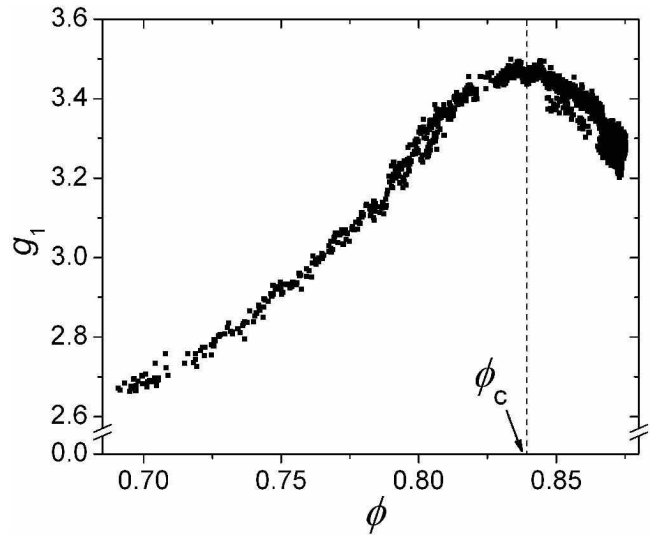}
\end{center}
\caption[Structural signature of the jamming transition]{Structural
signature of the jamming transition. The height of the first peak of
the pair correlation, $g_1$, is plotted as a function of $\phi$. The
position of the maximum is indicated as $\phi_c$.} \label{Figure6}
\end{figure}

To study the structure of the pack, we measure its pair correlation
function, $g(r)$ \cite{Zallen}.  As shown in Fig.~\ref{Figure5}, at
each $\phi$, $g(r)$ has an oscillating shape characteristic of any
disordered medium.  As $\phi$ increases from the unjammed phase, the
height of the first peak of $g(r)$, $g_1$, increases first. However,
when $\phi$ is above $\phi_c=0.84\pm0.02$, $g_1$ begins to decrease
(Fig.~\ref{Figure6}). Here, we measured the packing fractions from
the two-dimensional projection of images. Hence, the average size of
particles at $\phi_c$ can be estimated as $d=d_0
(\phi_c/\phi_\mathrm{initial})^{1/2}$, where $d_0=3.3$mm is the
average initial size of particles. We identify $\phi_c$ by fitting
$g_1(\phi)$ near its peak with a peak function. Thanks to symmetric
shape of $g_1(\phi)$ near $\phi_c$, a Gaussian function provides a
good fitting. Presumably any other similar peak functions would lead
to the same value of $\phi_c$. The non-monotonic trend in
$g_1(\phi)$ indicates a structural signature. Is this signature a
signature of jamming transition?  In other words, is $\phi_c$ the
jamming point?

\begin{figure}
\begin{center}
\includegraphics[width=3.35in]{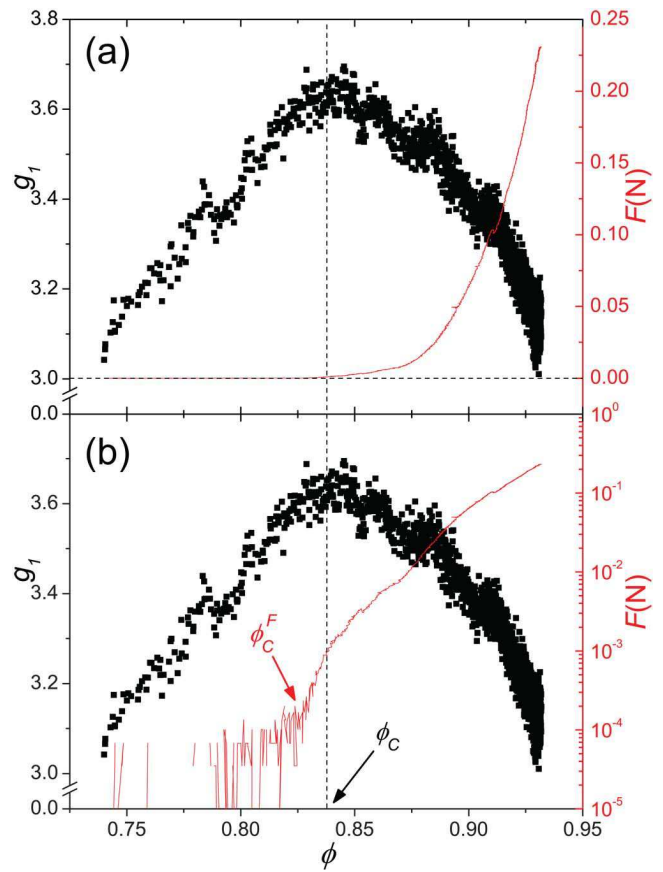}
\end{center}
\caption[Comparison of the structural signature $g_1(\phi)$ with the
force measured along the boundary $F(\phi)$]{(Color online)
Comparison of the structural signature $g_1(\phi)$ with the force
measured along the boundary $F(\phi)$ (a) in a linear-linear plot
and (b) in a logarithmic-linear plot. The black squares show $g_1(\phi)$ on
the left and the red line shows $F(\phi)$ on the right. The vertical
dashed line marks $\phi_c$ and the horizontal dashed line indicates
zero force. The onset of force $\phi_c^F$ is indicated in (b).}
\label{Figure7}
\end{figure}

To answer that, we checked two well-defined criteria for jamming.
First, the jamming point is supposed to mark the onset of rigidity
in a disordered system \cite{O'Hern}: a system composed of particles
with finite range purely repulsive interactions at $T=0$ begins to
build up pressure on its boundary at this point.  As shown in
Fig.~\ref{Figure7}a, the force measured at the boundary of the cell
is zero initially when the system is unjammed, and begins to deviate
from zero at a $\phi$ consistent with the peak of $g_1$. To
illustrate the detail of the onset of the jamming, we also plot the
force in the logarithmic scale (Fig.~\ref{Figure7}b). Below the
jamming point, the force fluctuates around the noise level (below
$100\mu$N) of the instrument. As $\phi$ increases further, the slope
of the curve changes sharply at $\phi=\phi_c^F$, which is indicated
by the left red arrow in Fig.~\ref{Figure7}b. Above $\phi_c^F$, the
pressure on the boundary increases significantly. Note that
$\phi_c^F$ is a little ahead of but very close to $\phi_c$. The
phenomenon is robust for experiments with uniform initial packing
fractions. We find that $\phi_c/\phi_c^F=1.009\pm0.006$, which
clearly suggests that, $\phi_c$, and therefore the structural
signature we found in $g_1(\phi)$, is directly related to the
jamming transition. We suggest that the small difference between
$\phi_c^F$ and $\phi_c$ is due to friction in the system. Particles
compressed onto the force sensor can be held in a force balance by
friction with other particles and with the bottom of the cell, and
therefore are not jammed globally with all the particles in the
system, which always results in a smaller $\phi_c^F$ than $\phi_c$.
However, the static friction in the aqueous system near the
isostatic point of the jamming transition is too small to sustain
much stress from the swelling of particles. The force balance
maintained by friction will break down quickly. Therefore,
$\phi_c-\phi_c^F\ll1$.

\begin{figure}
\begin{center}
\includegraphics[width=3.35in]{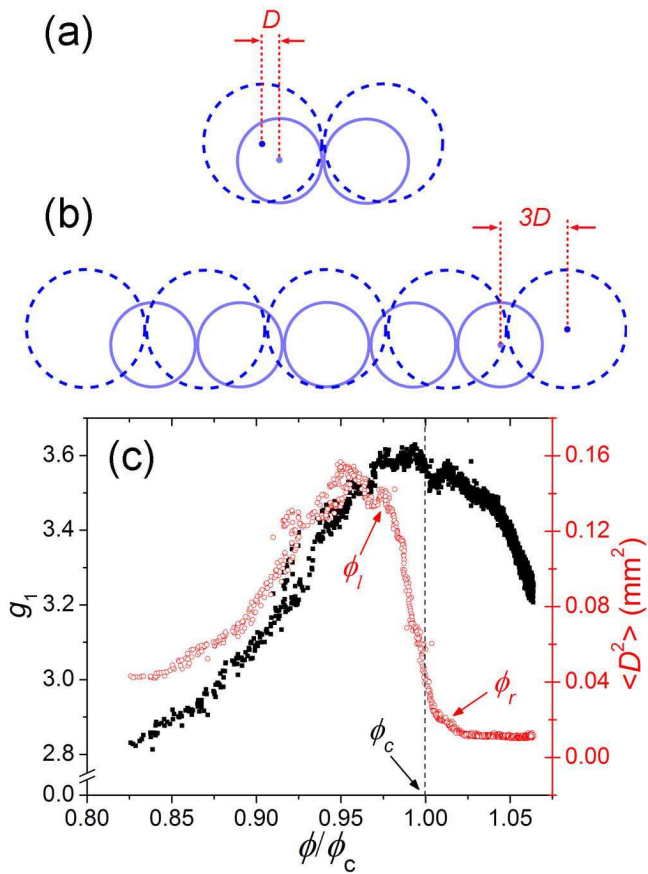}
\end{center}
\caption[Comparison of the structural signature $g_1(\phi)$ with the
mean square displacement $\langle D^2(\phi)\rangle$]{(Color online)
Comparison of the structural signature $g_1(\phi)$ with the mean
square displacement $\langle D^2(\phi)\rangle$. (a) and (b) are
sketches to show particle motion. If two particles touch (solid
circle), the centers of particles will be displaced after they swelled
up (dashed circle). (b) Particles sitting at the edge of a cluster
will be displaced most. The displacement is proportional to the size
of cluster. (c) Comparison between $g_1(\phi)$ and $\langle
D^2(\phi)\rangle$. The displacement of particles $D$ is measured in
a time interval $\Delta t=800$ s. The black squares show $g_1(\phi)$
on the left and the red circles show $\langle D^2(\phi)\rangle$ on
the right. The interval of decreasing $\langle D^2(\phi)\rangle$ is
indicated by $\phi_l$ on the left and by $\phi_r$ on the right. The
vertical dashed line indicates $\phi_c$. Due to random noise in the
system, $\langle D^2(\phi)\rangle$ does not approach zero at large
$\phi$. The source of the noise is discussed in the main text of
Sec. IV.} \label{Figure8}
\end{figure}

Another supporting evidence is from the motion of particles. Even
though the system is athermal, a particle can still be displaced
when it touches other particles during the swelling process
(Fig.~\ref{Figure8}a and \ref{Figure8}b). We shall discuss the displacement of
particles in more detail in Sec. IV.  Here, it is sufficient to
know that the motion of particles stops at the jamming point, which
is self-evident as the kinematic criterion of the jamming
transition. Ideally, in a homogeneous system, swelling particles
will touch the boundaries of the cell in different directions
simultaneously, which results in an extremely sharp drop of the
average velocity of particles at $\phi=\phi_c$. However, in an
inhomogeneous frictional system, some particles may reach a boundary
of the cell faster and stop the motion first, while particles in
other parts of the system still move. Therefore, there exists a
finite interval $\Delta\phi$ for diminishing of particle motion. For
an experiment with a uniform initial condition, the interval $\Delta
\phi$ is small. To show the average behavior of particles' motion
quantitatively, we measured the mean square displacement of
particles, $\langle D^2(\phi)\rangle \equiv \frac{1}{N} \sum_{i=1}^N
\big | \vec{D}_i(t(\phi)) \big |$, in our experiments. Here,
$\vec{D}_i(t(\phi))\equiv \vec{r}_i(t+\Delta t)-\vec{r}_i(t)$ is the
displacement of the particle $i$ at time $t$ within a small time
interval $\Delta t$ (or equivalently at packing fraction $\phi$
within an interval $\Delta \phi$ since the monotonic dependence of
$t(\phi)$) and the summation is over all the particles in the
studied area. $\vec{r}_i(t)$ is the center of the particle $i$ at
$t$. We fixed $\Delta t$ to be a small constant compared with the
time scale of the swelling of particles, so $\vec{D}_i(t)/\Delta t$
is approximately the instantaneous velocity of particle $i$ at $t$.
Hence, the mean squared velocity should show qualitatively the same
behavior as that of $\langle D^2 \rangle$. Note that the mean square
displacement defined here is different from the more common concept
used in studying the diffusion of Brownian particles. The result of
$\langle D^2(\phi)\rangle$ is shown in Fig.~\ref{Figure8}c. As one
can see from the plot, $\langle D^2(\phi)\rangle$ decreases to zero
quickly in an interval between $\phi_l$ and $\phi_r$ as indicated in
the figure. For all experiments with uniform initial packing
fraction, we found that $\Delta \phi/\phi_c \equiv
(\phi_r-\phi_l)/\phi_c=0.03\pm0.01$. More importantly, $\phi_c$ is
always located between $\phi_l$ and $\phi_r$ (i.e.
$\phi_l<\phi_c<\phi_r$).  This confirms the argument that the
structural signature in $g_1(\phi)$ is due to the jamming
transition.

Both the mechanical and kinematic measurements show that $\phi_c$
indeed corresponds to the jamming point of frictional systems.
Therefore, the peak in $g_1(\phi)$ manifests the structural
signature of the zero-temperature jamming transition. But how can we
understand this structural signature? What is the physical or
geometric origin of it? We shall discuss it next.

\subsection{B. Discussion}

It is well known that when a molecular liquid goes through glass
transition upon decreasing $T$ in the jamming phase diagram, no
structural signature can be observed \cite{Ediger,Busse}; $g_1$
monotonically increases as the system passes through the glass
transition temperature $T_g$.  Then why does $g_1$ show a peak when
the system goes through jamming transition along the $1/\phi$ axis?
This was discussed in Ref.~\cite{Zhang}, where a vestige of the
zero-temperature signature was observed in a finite temperature
colloidal system.  When $\phi$ approaches $\phi_c$ from the unjammed
side, particles are pushed closer to each other.  Therefore, $g_1$,
which indicates the probability that nearest-neighbors of a particle
are located at the same distance, increases as the total number of
nearby neighbors increases.  However, above the jamming transition,
particles begin to overlap and deform.  Since the degree of
deformation depends on the local environment of a particle, the
distribution of distances between two particles in contact becomes
broader.  Meanwhile, the number of nearest-neighbor particles (the
coordination number) does not increase appreciably, {\it i.e.}, the
area under the first peak of $g(r)$ remains roughly constant.  As a
result, $g_1$, the height of the first peak in $g(r)$,  decreases.
In simulations with monodisperse frictionless particles, at the
jamming point, all particles are precisely one particle diameter
away from their nearest neighbors.  Hence, $g_1$ diverges at this
point \cite{Silbert2}.  In our system, the polydispersity of the
particles reduces $g_1$.  Even without friction, at the jamming
point, the distribution of distance between two particles in contact
still has a finite width reflecting the size distribution.

Friction may complicate the situation further. Both the coordination
number and the deformation of particles under compression are
profoundly changed in the presence of friction as indicated in
previous simulation and theoretical works
\cite{Makse,Silbert3,Shundyak,Somfai}. Hence, it is not
straightforward to extend the results of ideal frictionless system
on which most simulation and theoretical studies focus
\cite{O'Hern,Silbert1,Silbert2,Ellenbroek,Zeravcic,Henkes,Wyart1,Wyart2}
to experiments with real frictional granular matter. Experimentally,
Majmudar {\it et al.} found that the increasing of the coordination
number, $Z-Z_c$, and the pressure, $P$, of the system as a function
of $\phi-\phi_c$ agrees with the mean-field theory for frictionless
particles \cite{Majmudar}. However, the two experiments on the sound
propagation near the surface of loosely compacted granular packs
show that the ratio of the shear modulus to the bulk modulus, $G/B$,
stays constant rather than diminishing as the pressure of the system
approaches zero \cite{Jacob,Clement}, which contradicts to the
result of frictionless particles \cite{O'Hern,Wyart2}. Until now, no
direct measurement has been conducted on the structural signature of
the jamming transition in a real granular system. Here, we show that
structural signature predicted with frictionless particles
\cite{Silbert2} persists in the system of real granular matter with
frictional contact, although the signature is modified significantly.

\begin{figure}
\begin{center}
\includegraphics[width=3.35in]{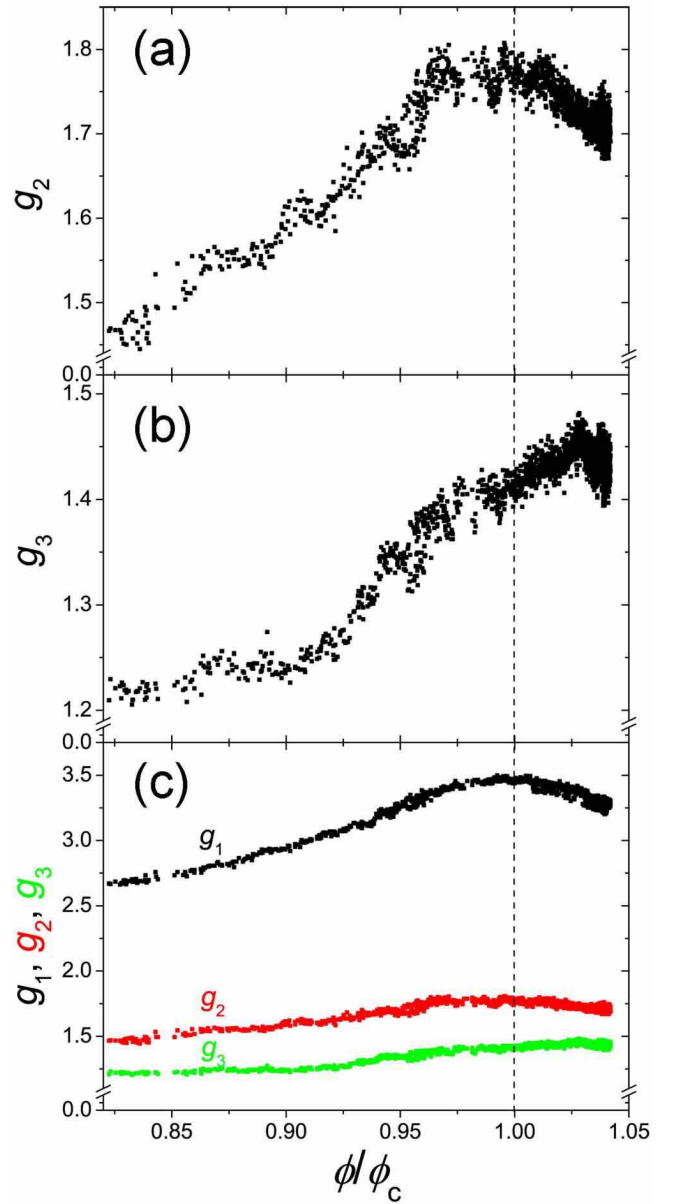}
\end{center}
\caption[Structural signature in higher order peaks of the pair
correlation function]{(Color online) Structural signature in higher
order peaks of the pair correlation function. (a) Height of the
second peak of pair correlation function, $g_2$, as a function of
$\phi$. (b) Height of the third peak of pair correlation function,
$g_3$, as a function of $\phi$. (c) Comparison of the amplitudes of
$g_1(\phi)$, $g_2(\phi)$ and $g_3(\phi)$. The vertical dashed line
indicates $\phi/\phi_c=1$.} \label{Figure9}
\end{figure}

Finally, it is also interesting to look for this structural
signature in the other peaks of the pair correlation function. The
system with the larger cell contains enough particles to show the
first four peaks of $g(r)$ clearly (Fig.~\ref{Figure5}). As seen in
Fig.~\ref{Figure9}a, the height of the second peak of $g(r)$, $g_2$,
as a function of $\phi$ also shows a maximum at the same value of
$\phi_c$, at which $g_1(\phi)$ shows a maximum.  However, the
amplitude of this maximum is smaller than that of $g_1$
(Fig.~\ref{Figure9}c).  One may think that the height of the $n$th
peak of the pair correlation function, $g_n$, would also show a peak
at $\phi_c$ but with decreasing amplitude as $n$ increases. However,
when we measured the height of the third peak, $g_3(\phi)$, no peak
is found: instead $g_3$ increases rapidly at small $\phi$ and slows
down or plateaus in some cases at large $\phi$
(Fig.~\ref{Figure9}b).  The structural signature of jamming
transition apparently manifests itself in the first and the second
peaks of the pair correlation function but not in the peaks at
larger separation. As a comparison, in the simulations with
frictionless monodisperse particles, the second peak of the pair
correlation function splits into two sub-peaks
\cite{Silbert2,Zallen} both of which have a divergent slope
\cite{Silbert2}. Furthermore, no singular behavior of $g_3$ is found
in simulations \cite{Silbert2}. So far, unlike the case of $g_1$, no
clear geometric picture exists for why $g_2$ should diverge at the
jamming point \cite{Silbert2}. It is hypothesized that the
underlying mechanism for the singularity of simulations may be
related to the experimental finding.

\section{IV. Formation of static cluster structure}

In this section, we investigate the displacement field of the
particles.  We shall show that, for any typical pack, the system
eventually organizes itself into clusters.  One can extract a static
length scale from the sizes of the clusters.  This length grows
dramatically from the size of a few particles to the size of the
entire system when the system approaches the jamming point.

\subsection{A. Displacement of particles}

The centers of two contact particles separate due to the enlargement
of their radiuses as illustrated in Fig.~\ref{Figure8}a. Thus, by
tracking centers of particles, one can observe clear motion in this
athermal system.

Now let us first have a look at the average behavior of particles at
different stages through the jamming transition. As shown in
Fig.~\ref{Figure8}c, the mean square displacement of particles,
$\langle D^2(\phi)\rangle$, shows a non-monotonic behavior. Since
only a few pairs of particles have contacts initially, $\langle
D^2(\phi)\rangle$ is small at the beginning (Fig.~\ref{Figure8}c).
As the size of particles increases, more particles form contacts and
are displaced; $\langle D^2(\phi)\rangle$ increases correspondingly.
When the system approaches the jamming point, almost all the
particles join into a contact network so that $\langle
D^2(\phi)\rangle$ reaches a maximum.  However, shortly after that,
the moving particles begin to touch the boundary and $\langle
D^2(\phi)\rangle$ quickly drops toward zero as the entire system
jams. The position of the peak $\langle D^2(\phi)\rangle$ is always
before the peak of $g_1(\phi)$. Deep inside the jammed regime, there
are rare buckling events: particles suddenly change their relative
positions on a much shorter time scale than that for particle
swelling. These events are discrete and localized --- a typical
buckling event involves only two to five particles. This is reminiscent
of the T1 process found in a two-dimensional foam \cite{Weaire}.

\subsection{B. Displacement field and cluster structure}

\begin{figure*}[t]
\begin{center}
\includegraphics[width=6in]{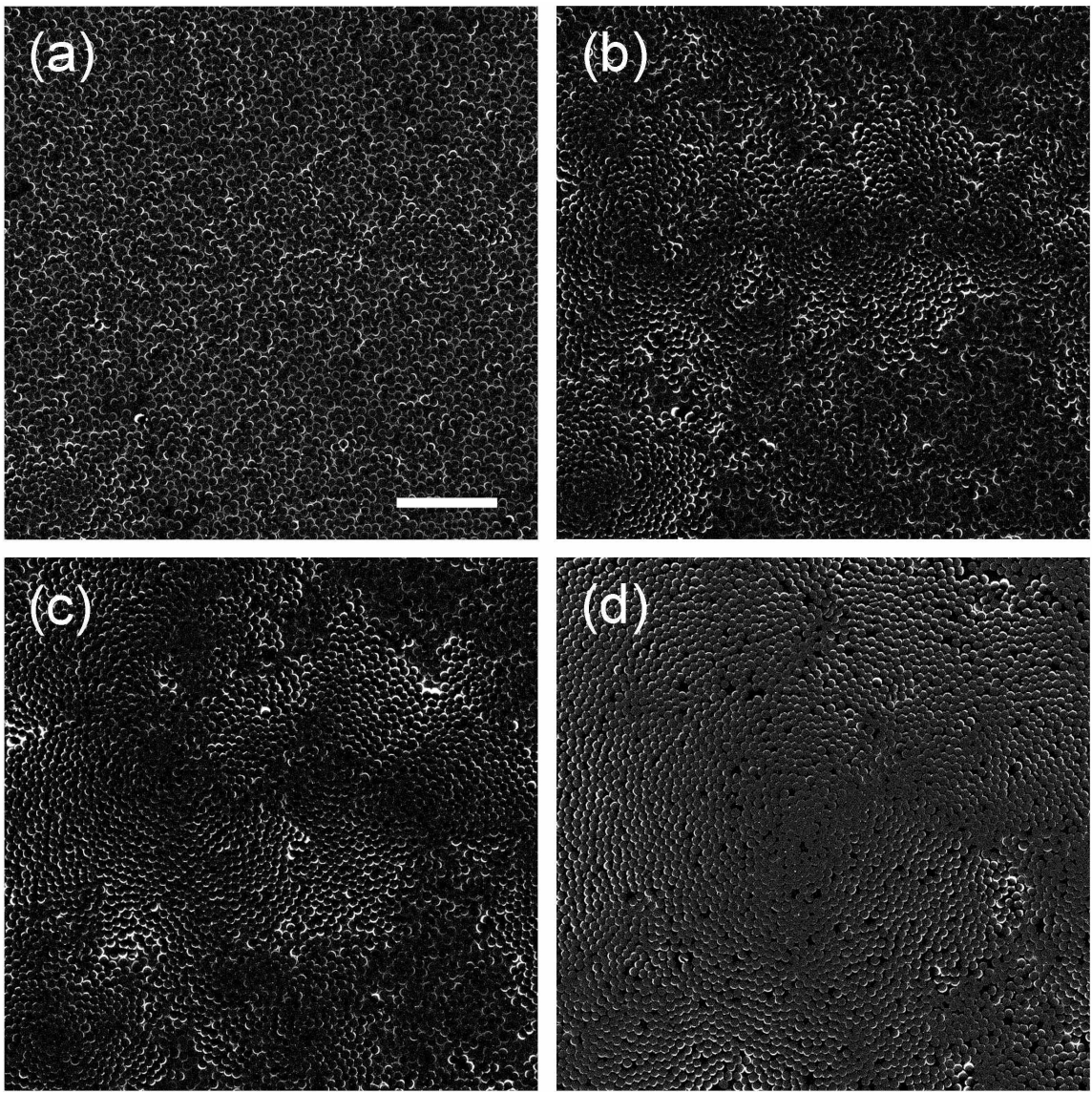}
\end{center}
\caption[Cluster structure]{Cluster structure. Displacement field of
a subsystem in the large cell (54.6 cm $\times$ 54.6 cm). The time
interval for the displacement is $\Delta t=800$ s. (a) $\phi=0.65$
at $t=0$ h, (b) $\phi=0.74$ at $t=1.30$ h, (c) $\phi=0.79$ at
$t=2.29$ h, and (d) $\phi=0.84$ at $t=3.92$ h. The white scale bar is 5
cm.} \label{Figure10}
\end{figure*}

More information can be obtained from the displacement field of the
system. To visually illustrate the particle-displacement field, we
subtract two images of the system at different times.  Any
stationary part of the system will appear black in such difference
images since the individual pixels are identical in that region.
However, if a particle moves, the image subtraction will produce an
area of crescent shape with positive values along the front boundary
in the moving direction and leave a similar crescent area with
negative values in the rear. Furthermore, we assign the negative
values to zero (black). Hence, only front boundaries in the moving
direction are indicated in the image. The curvature and area of the
crescent show the direction and magnitude of a particle's
displacement, respectively (Fig.~\ref{Figure10}a-\ref{Figure10}d).

Fig.~\ref{Figure10}a-\ref{Figure10}d show the displacement field of a typical
experiment. The displacement field is initially random
(Fig.~\ref{Figure10}a). As the system evolves toward higher packing
fractions, a coherent structure emerges (Fig.~\ref{Figure10}b and \ref{Figure10}c).
Particles tend to move outward around a few nuclei. In other words,
a few clusters form in the system. Eventually, when the system
approaches the jamming point, one single cluster forms
(Fig.~\ref{Figure10}d).

How does the cluster structure emerge out of a random initial
configuration? As mentioned above, since only a few pair of
particles are in contact initially, the average displacement is
small and the directions of displacement are random at the beginning
(Fig.~\ref{Figure10}a). As the size of particles increases, they
begin to form local contact networks --- cluster structure emerges
(Fig.~\ref{Figure10}b). Encircled by its neighbors, a particle at
the center of a cluster feels zero average force due to the balance
of the compression from different directions. Hence, it stays in
stationary and shows as a dark nucleus in the displacement field
(Fig.~\ref{Figure10}b). Meanwhile, the surrounding particles move
away from the center radially. Due to inevitable initial density
variations, the denser part of system will form clusters first.
Since the displacement of particles at the edge of a cluster is
linearly proportional to the size of the cluster
(Fig.~\ref{Figure8}a and \ref{Figure8}b), a larger cluster will grow faster, and
therefore aggregate more particles.  Thus, the clusters formed at
early time in the denser part of system will quickly dominate the
system and the small initial density variation is amplified.  When
two clusters meet, the internal particles and especially those along
the cluster boundaries rearrange and the two clusters merge into a
single bigger entity (Fig.~\ref{Figure10}b and \ref{Figure10}c). This merging of
clusters continues until a single system-spanning cluster forms when
the system approaches the jamming point (Fig.~\ref{Figure10}d).

The exact shape of evolving clusters depends on the initial
condition of experiments.  For the experiment shown in
Fig.~\ref{Figure10}, the initial density is higher in the central
area of the system.  Hence, the cluster structures appear first in
that area and then expand outwardly.  Although details depend on a
packing's history, the existence and development of cluster
structures are robust.  Unless over 10,000 particles in the system
start out with, and maintain, exactly the same inter-particle
spacings, which clearly is unrealistic, the system will always
evolve into cluster structures at later time.  Any small initial
density variation will be amplified as the system approaches the jamming
point.

\subsection{C. Quantitative analysis}

\begin{figure}
\begin{center}
\includegraphics[width=3.35in]{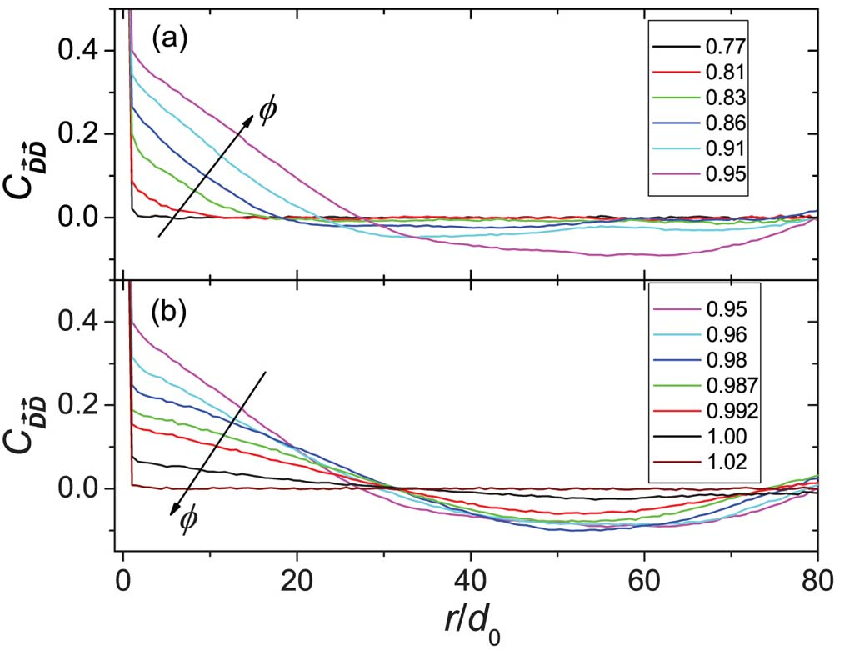}
\end{center}
\caption[Correlation of the displacement of particles,
$C_{\vec{D}\vec{D}}(r)$, (a) before the mean square displacement reaches
the maximum and (b) after the maximum near the jamming point
(b)]{(Color online) Correlation of the displacement of particles,
$C_{\vec{D}\vec{D}}(r)$, before the mean square displacement reaches
the maximum (a) and after the maximum near the jamming point (b).
The arrows indicate the direction of increasing packing fraction.
Values of $\phi/\phi_c$ are shown in the plots. The time interval
for the displacement is $\Delta t=200$ s.} \label{Figure11}
\end{figure}

\begin{figure}
\begin{center}
\includegraphics[width=3.35in]{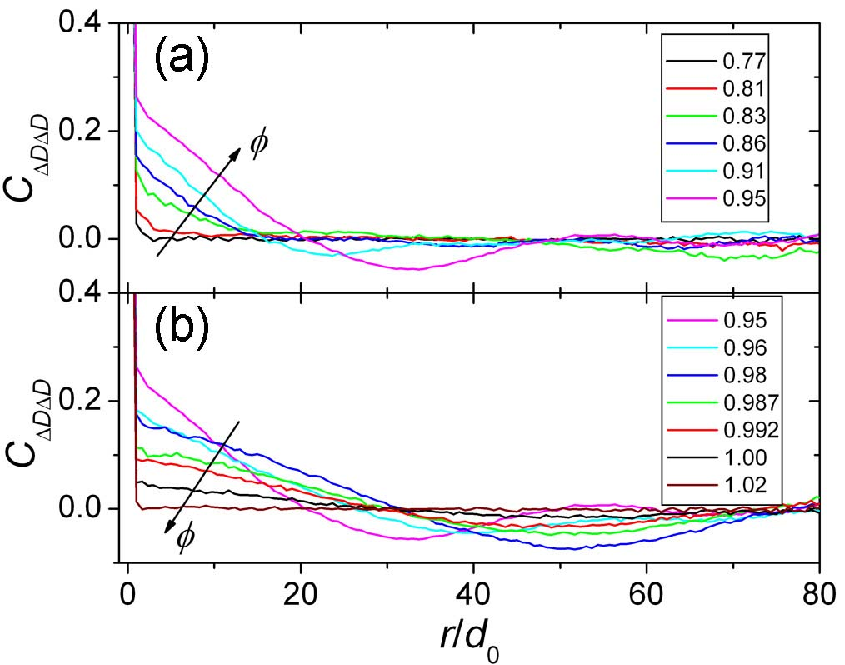}
\end{center}
\caption[Correlation of the magnitude of particle displacement,
$C_{\Delta D\Delta D}(r)$, before the mean square displacement
reaches the maximum (a) and after the maximum near the jamming point
(b)]{(Color online) Correlation of the magnitude of particle
displacement, $C_{\Delta D\Delta D}(r)$, (a) before the mean square
displacement reaches the maximum and (b) after the maximum near the
jamming point (b). The arrows indicate the direction of increasing
packing fraction. Values of $\phi/\phi_c$ are shown in the plots.
The time interval for the displacement is $\Delta t=200$ s.}
\label{Figure12}
\end{figure}

To quantify the cluster structure, we measure the two-point
correlation function of the displacement field,
\begin{equation}
\label{CDD} {C_{\vec{D}\vec{D}}(r) = \frac{\displaystyle{
\frac{1}{N_0} \sum_{i,j=1}^N \left(
\vec{D}(\vec{r}_i)\cdot\vec{D}(\vec{r}_j) \right) \delta(r_{ij}-r)}}
{\displaystyle{\frac{1}{N}\sum_{i=1}^N\vec{D}(\vec{r}_i)\cdot\vec{D}(\vec{r}_i)}}},
\end{equation}
which is usually used to identify coherent structures in a system.
Here, $\vec{D}(\vec{r}_i)$ is the displacement of the particle $i$
located at $\vec{r}_i$ within a small time interval $\Delta t$, and
$r_{ij}=|\vec{r}_i-\vec{r}_j|$. The summations are over all the $N$
particles in the system and the normalization factor
$N_0={\sum_{i,j=1}^N\delta(r_{ij}-r)}$. Experimentally, the data are
binned with a bin size of two thirds of a particle diameter. As one
can see in Fig.~\ref{Figure11}a, initially at low packing fraction
($\phi/\phi_c=0.77$) no correlation exists,
$C_{\vec{D}\vec{D}}(r>0)=0$.  As the packing fraction increases,
clusters begin to form. Correspondingly, both the correlation and
the correlation length increase (Fig.~\ref{Figure11}a). For example,
the point where $C_{\vec{D}\vec{D}}$ crosses zero shifts to larger
$r$ as $\phi$ increases. The correlation reaches a maximum when the
mean square displacement $\langle D^2(\phi)\rangle$ is largest.  As
$\langle D^2(\phi)\rangle$ falls near the jamming point
(Fig.~\ref{Figure8}c), the correlation magnitude decreases but the
length scale of the correlations is fixed at the size of the system
(Fig.~\ref{Figure11}b). One can of course choose other correlation
functions for identifying cluster structures.  The correlation
function chosen here includes the information of both the direction
and magnitude of displacements.  We also measure the correlation of
only the magnitudes of particle displacement, which is defined as
\begin{equation}
\label{CdeltaDdeltaD} {C_{\Delta D\Delta D}(r)=
\frac{\displaystyle{\frac{1}{N_0}\sum_{i,j=1}^N\Delta
D(\vec{r}_i)\Delta D(\vec{r}_j)\delta(r_{ij}-r)}} {\displaystyle{
\frac{1}{N}\sum_{i=1}^N\Delta D(\vec{r}_i)\Delta D(\vec{r}_i)}}},
\end{equation}
where $\Delta D(\vec{r}_i)=D(\vec{r}_i)-\langle D \rangle$ is the
fluctuation of the magnitude of particle displacement at
$\vec{r}_i$. $D(\vec{r}_i)$ is the magnitude of particle
displacement, and $\langle D \rangle$ is the average magnitude of
particle displacement. We show $C_{\Delta D\Delta D}(r)$ at
different packing fractions in Fig.~\ref{Figure12}. As one can see,
$C_{\Delta D\Delta D}(r)$ shows qualitatively the same behavior as
$C_{\vec{D}\vec{D}}(r)$.

$C_{\vec{D}\vec{D}}(r)$ (or $C_{\Delta D\Delta D}(r)$) shows a clear
trend of the increasing of correlation between the displacements of
particles. It provides a good quantitative illustration of the
emergence and the evolution of cluster structures. However, it is
hard to extract the length scale of clusters from this function
directly --- $C_{\vec{D}\vec{D}}(r)$ does not decrease exponentially
with $r$ and there is no obvious feature in $C_{\vec{D}\vec{D}}(r)$.
Furthermore, due to the random noise in the displacement of
individual particles, the correlation function shows a sharp jump
from $C_{\vec{D}\vec{D}}(0)=1$ to $C_{\vec{D}\vec{D}}(r=d_0)$.  A
particle always perfectly correlates with itself, but the
correlation between different particles is reduced due to random
noise.  Hence, there exists a sharp jump from the self-correlation
at $r=0$ to the correlation of neighboring particles at $r=d_0$. The
jump is more severe when the signal/noise ratio is smaller at the
beginning of the experiment or in the jammed phase
(Fig.~\ref{Figure11}).  (At the beginning, when particles do not
touch and at the end when particles are jammed, the signal/noise ratio is
$\sim0$.)  The random noise can be due to the non-uniform swelling
of particles.  If the shape of a particle changes during swelling,
the center of the particle may move slightly even without contact
with other particles.  The particle-tracking algorithm can also
induce some noise.  However, that only happens when the system is
deep inside the jammed phase where the inter-particle spacing is so
small that it becomes hard to distinguish the boundary between two
neighboring particles.

\begin{figure}
\begin{center}
\includegraphics[width=3in]{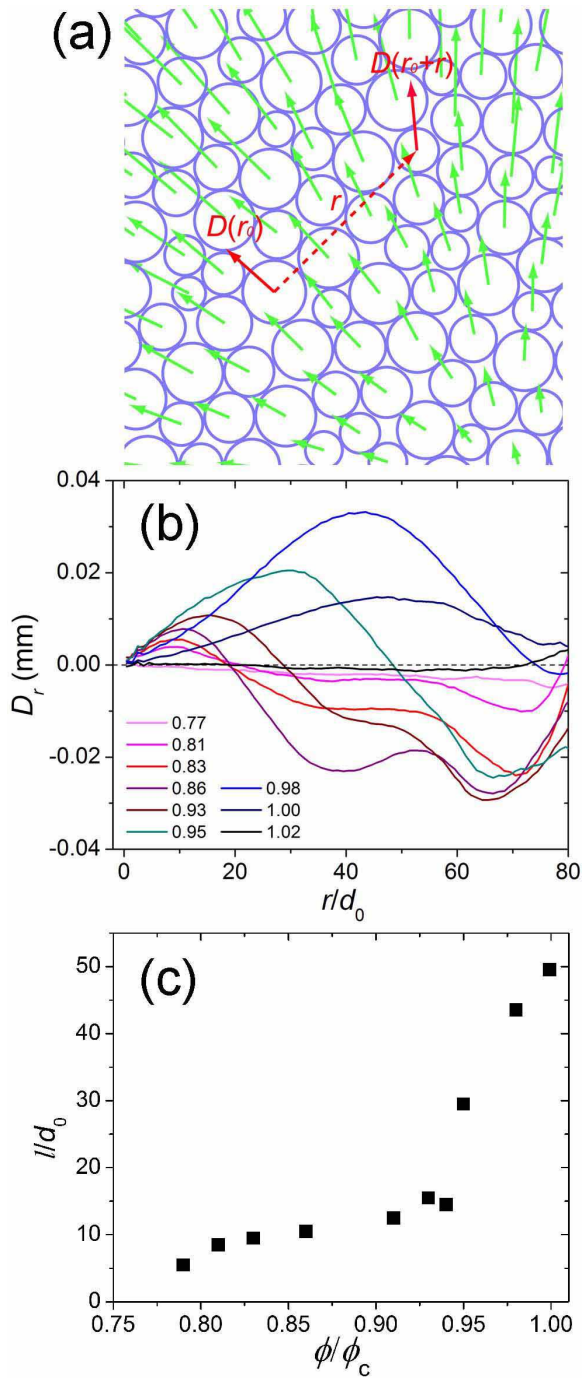}
\end{center}
\caption[Projection of the relative displacement of two particles on
the direction of their relative position, $D_r$, and length scale,
$l$, extracted from the cluster structure]{(Color online)
Projection of the relative displacement of two particles on the
direction of their relative position, $D_r$, and length scale, $l$,
extracted from the cluster structure. (a) Illustration of the
definition of $D_r$. A pair of particles with distance $r$ (shown
with red arrows) always move apart inside a cluster. (b) $D_r$ as a
function of $r$ for different packing fractions. The time interval
for the displacement is $\Delta t=200$ s. Values of $\phi/\phi_c$
are shown in the plot. The horizontal dashed line indicates $D_r=0$.
(c) Length scale, $l$, extracted from $D_r$, as a function of
$\phi/\phi_c$.} \label{Figure13}
\end{figure}

Another way to quantify the cluster formation is to measure the
projection of the relative displacement of two particles on the
direction of their relative position (Fig.~\ref{Figure13}a):
\begin{equation}
\label{Dr} {D_r(r)=
\frac{1}{N_0}\sum_{i,j=1}^N\left(\left(\vec{D}(\vec{r}_i)-\vec{D}(\vec{r}_j)\right)\cdot\frac{\vec{r}_{ij}}{r_{ij}}\right)\delta(r_{ij}-r)}.
\end{equation}
The bin size is again chosen as two thirds of a particle diameter.
$D_r(r)$ indicates on average whether a pair of particles at
separation $r$ moves closer $(D_r(r)<0)$ or moves apart
$(D_r(r)>0)$.  If there are no clusters and all the particles move
randomly, $D_r(r)$ will be zero.  However, inside a cluster, any two
particles cannot move closer (Fig.~\ref{Figure13}a).  Therefore, in a
cluster with size $r_0$, $D_r(r<r_0)\ge0$. When all particles are
stationary, $D_r(r)=0$. Figure~\ref{Figure13}b shows $D_r(r)$ at
different packing fractions.  As one can see, initially there is no
coherent motion and $D_r(r)$ is flat at zero.  However, as the
system evolves, $D_r(r)$ begins to deviate from zero and a region
with a positive $D_r(r)$ clearly shows up implying the presence of
clusters.  Although the amplitude of this positive region shows a
non-monotonic behavior, the position of the peak at $l$ increases
with $\phi$.  Thus, $l$ can be used as a characteristic length scale
of clusters.  As shown in Fig.~\ref{Figure13}c, $l$ increases slowly
from a few particles diameter at low $\phi$ and increases rapidly to
the size of the system when the system approaches the jamming point.
The exact shape of $l(\phi)$ depends on the initial configuration of
the particles, but qualitatively all packs show the same behavior.

\subsection{D. Discussion}

In contrast to the length scales determined from dynamic
heterogeneities in supercooled liquids \cite{Ediger2,Berthier},
colloidal suspensions \cite{Weeks,Kegel} and granular media
\cite{Dauchot,Lechenault,Keys}, the length measured here reflects
the static structure of system.  One might imagine that the average
packing fraction inside a cluster is higher than that outside the
cluster. The cluster structure observed here is due to the athermal
nature of the system. Clearly, the cluster structure depends on the
initial packing configuration. Although we prepare packs in a random
way with unavoidable small density variations, the system always
amplifies this initial variation into a cluster structure at later
time.  Such a history dependence and memory are typical of
non-equilibrium systems \cite{Jonason,Angell,Josserand}. On the
contrary, in any equilibrium system without attraction, the initial
density fluctuations will quickly be smeared out; unless the system
approaches a second order phase transition, the correlations of the
density fluctuations in equilibrium systems will remain small. The
existence of a divergent static length scale in the glass or jamming
transitions would be a hallmark for an underlying phase transition.
However, until now, no unambiguous static length scale has yet been
observed in any equilibrium system near the glass or jamming
transition.  The present experiment suggests that at the jamming
transition a {\it non-equilibrium} system can produce a divergent
static length scale. It should be noted that previous simulation
works on the jamming transition at zero temperature acquire the
unjammed configuration by quenching the system from $T=\infty$ to
$T=0$ \cite{O'Hern,Silbert1,Silbert2,Ellenbroek,Somfai}. Therefore,
even though the system is at $T=0$, the static configuration of the
system is essentially the same as that at $T=\infty$. Contacts
between particles do not exist before jamming and static length
scale cannot be found in these simulations.

\section{V. Effect of friction: Multiple jamming points}

As shown in the above section, the initial density variation, no
matter how small it is, will be amplified by the system {\it en
route} to the jammed phase and the system will spontaneously
organize into cluster structure. In this section, we shall
investigate how this density variation influences the signature of
the jamming transition. We shall show that for a system with
sufficient large initial density variation, {\it i.e.} for a highly
inhomogeneous system, friction plays an important role and multiple
jamming points exist.

Practically, the initial density variation can be effectively
controlled by the total number of particles in the system. If enough
particles are added at the beginning, the initial packing fraction
of system will be high and the density variation will be small.  In
this case, although the cluster structure still emerges while the
system approaches the jamming point, the system will eventually show
a clear jamming transition and therefore an unambiguous structural
signature of the transition as shown in Sec. III. However, if the
sample is prepared at a low initial packing fraction, due to random
vibrations during the sample preparation, there is a good chance
that one part of the system is denser than the rest.  The initial
density variation of the system will be much higher. Hence, we can
easily prepare a highly inhomogeneous system by simply reducing the
initial packing fraction.  With a highly inhomogeneous system at
hand, we want to ask how such a system goes through the jamming
transition. Is the jamming signature of a highly inhomogeneous
system the same as that of a homogeneous system?

\begin{figure}
\begin{center}
\includegraphics[width=3.35in]{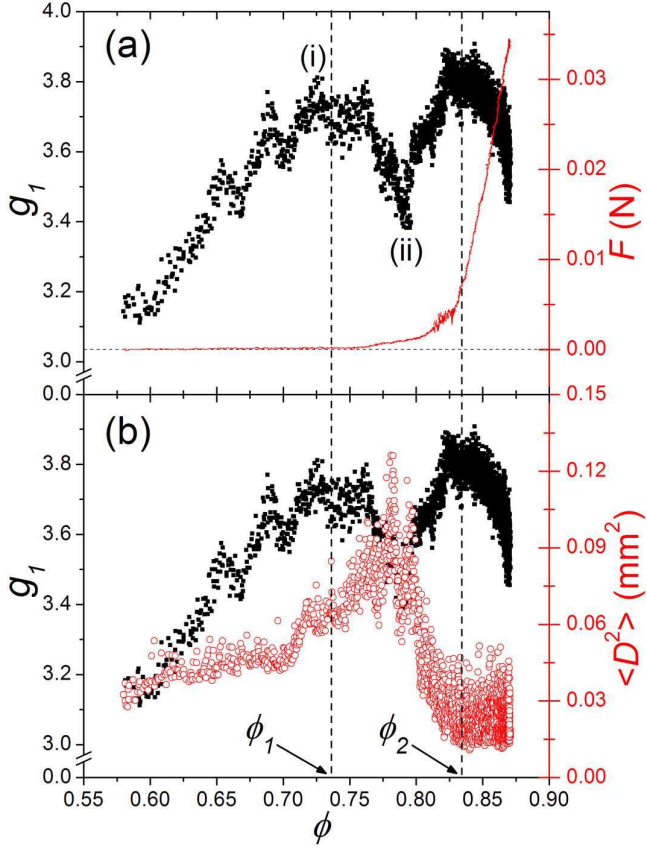}
\end{center}
\caption[Height of the first peak of the pair correlation function,
$g_1$, as a function of $\phi$ for a system with low initial packing
fraction, and comparison of $g_1(\phi)$ with (a) the force along the
boundary $F(\phi)$, and with (b) the mean square displacement of
particles $\langle D^2(\phi)\rangle$]{(Color online) Height of the
first peak of the pair correlation function, $g_1$, as a function of
$\phi$ for a system with low initial packing fraction, and
comparison of $g_1(\phi)$ with (a) the force along the boundary
$F(\phi)$, and with (b) the mean square displacement of particles
$\langle D^2(\phi)\rangle$. $g_1(\phi)$ is shown on the left with
black squares. The vertical lines mark the positions of two major
peaks, $\phi_1$ and $\phi_2$. (i) and (ii) indicates the positions
where we show the displacement field in Fig.~\ref{Figure15}.
$F(\phi)$ is shown on the right of (a) with red line. The horizontal
dashed line indicates zero force. $\langle D^2(\phi)\rangle$ is
shown on the right of (b) with red circles. The time interval for
the displacement is $\Delta t=800$ s.} \label{Figure14}
\end{figure}

Figure~\ref{Figure14}a shows the height of the first peaks of the pair
correlation function, $g_1$, as a function of $\phi$ for a system
with low initial packing fraction (typically
$\phi_\mathrm{initial}\leq0.60$). Instead of a single pronounced
peak, there are two major peaks located at $\phi_1\simeq0.74$ and
$\phi_2\simeq0.84$ (and possible several smaller peaks).  We can
compare $g_1(\phi)$ with the mechanical and kinematic criteria of
the jamming transition separately.  As shown in
Fig.~\ref{Figure14}a, the pressure on the boundary begins to deviate
from zero near the first peak $\phi_1$. However, if one compares
$g_1(\phi)$ with the mean square displacement, $\langle D^2\rangle$,
one finds that $\langle D^2\rangle$ drops to zero at the second peak
$\phi_2$ (Fig.~\ref{Figure14}b).  Therefore, the system jams at
$\phi_1$ according to the mechanical criterion for the jamming
transition and jams at $\phi_2$ according to the kinematic
criterion.  Why do the two well-defined criteria for the jamming
transition occur at two different packing fractions?

\begin{figure*}[t]
\begin{center}
\includegraphics[width=6in]{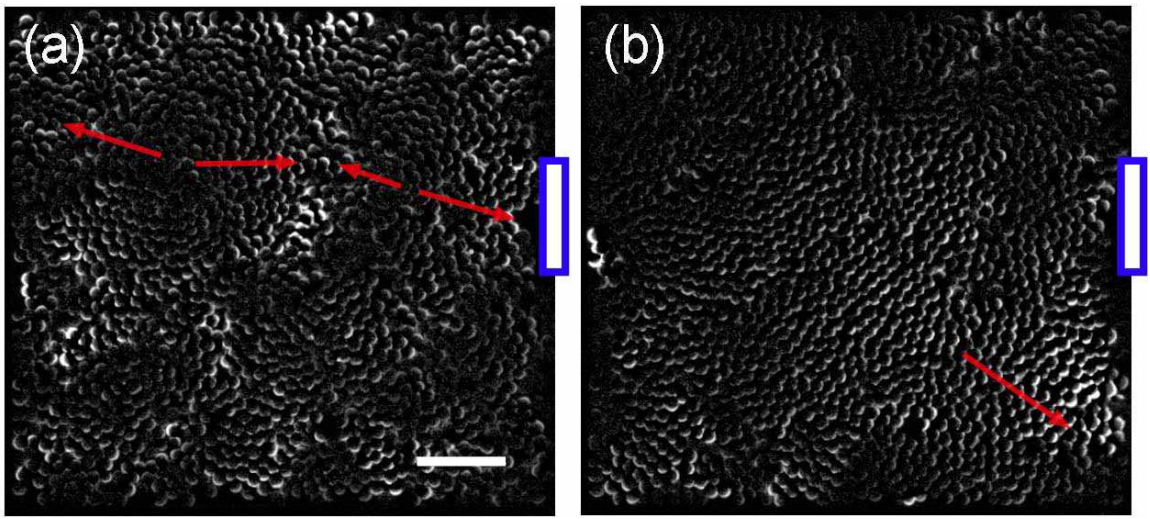}
\end{center}
\caption[Displacement field of an inhomogeneous system in the
smaller cell (16.5 cm $\times$ 16.5 cm)]{(Color online)
Displacement field of an inhomogeneous system in the smaller cell
(16.5 cm $\times$ 16.5 cm). (a) Displacement field near the first
peak of $g_1(\phi)$ at point (i) shown in Fig.~\ref{Figure14}. (b)
Displacement field at the valley between the first and the second
peaks at point (ii) shown in Fig.~\ref{Figure14}. The time interval
for the displacement is $\Delta t=800$ s. Red arrows show the local
mean displacement. The blue box on the right indicates the force
sensor at the boundary of the cell. The white scale bar is 3 cm.}
\label{Figure15}
\end{figure*}

To understand this, we plot the displacement field at two different
packing fractions in Fig.~\ref{Figure15}: (i) at the first peak
$\phi_1$ (Fig.~\ref{Figure14}a), and (ii) at the valley between
$\phi_1$ and $\phi_2$ (Fig.~\ref{Figure14}a).  As one can see, at
$\phi_1$ the system already forms clusters (Fig.~\ref{Figure15}a).
From the local mean displacement (red arrows), one can see that the
two clusters span the system from left to right.  The cluster on the
right pushes onto the force sensor at the boundary of cell. Clearly,
the system jams and a force chain forms along this direction from
left to right. However, there is still empty space in the lower
right corner of cell.  After the two clusters merge, the particles
move together towards the lower right corner as shown in the
displacement field near the valley (Fig.~\ref{Figure15}b).  The
system jams globally at $\phi_2$, which results in the dramatic drop
of the mean square displacement. Therefore, the reason why there are
two peaks in $g_1(\phi)$ is that the system jams in two steps: it
first jams locally along certain direction (from left to right) and
then it jams globally. For any system like this, depending on where
the force sensor is located, the pressure measured may deviate from
zero at different packing fractions.  But it should always happen
before the global jamming point ($\phi_2$), as confirmed by the
experiments.

With frictionless particles, the system can only jam as a whole due
to force balance throughout the entire sample.  Therefore, the
multiple jamming points found in the system have to be due to
friction between particles and between the particles and the
boundary of cell.   A frictional granular system can form very
inhomogeneous structures such as force chains along one specific
direction and jam in that direction, but still have empty space in
other directions.  As the size of particles increases further, force
chains will buckle under the increased stress, displacing particles
into the less dense regions adjacent to the chains, and causing the
system to unjam.  As a result, the system approaches the transition
to jamming in a series of steps.  This picture is reminiscent of the
scenario proposed based on theoretical considerations that rigidity
emerges by successive buckling of force chains in glasses and
granular matter \cite{Cates,Alexander}. It should be emphasized that
frictional contact is an essential ingredient for the existence of
force chains before the global jamming point. However, deep inside
the jammed phase, where the rigidity is already well established,
force chains can sustain without friction. In conclusion, with
friction the picture of a single jamming transition at $T=0$
(Fig.~\ref{Figure1}) has to be modified.  Multiple jamming points
may exist in a frictional system when a highly inhomogeneous
structure is present.

\begin{figure}
\begin{center}
\includegraphics[width=3.35in]{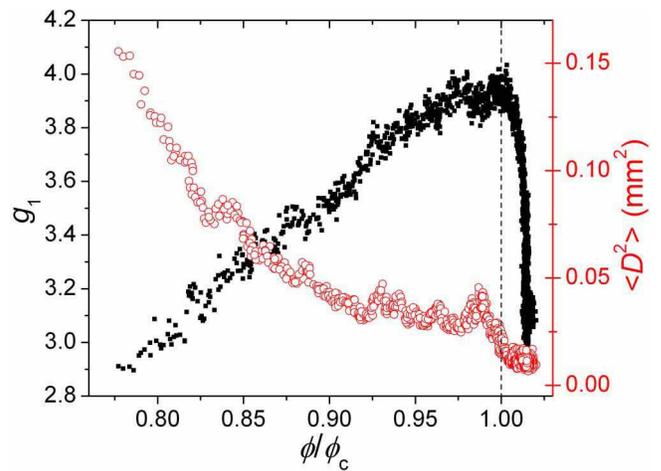}
\end{center}
\caption[Structural signature of the jamming transition of a system
subjected to small amplitude vibrations]{(Color online) Structural
signature of the jamming transition of a system subjected to small
amplitude vibrations. The height of the first peak of pair
correlation function $g_1$ is shown on the left with black squares.
The mean square displacement of particles $\langle D^2(\phi)\rangle$
is shown on the right with red circles. The time interval for the
displacement is $\Delta t=200$ s. $\phi$ is normalized by the
packing fraction at the jamming point $\phi_c\simeq0.83$, which is
indicated by the vertical dashed line.} \label{Figure16}
\end{figure}

\begin{figure}
\begin{center}
\includegraphics[width=3.35in]{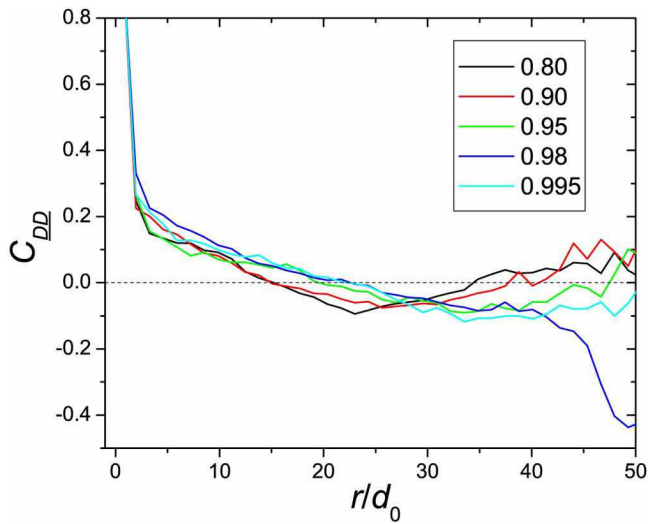}
\end{center}
\caption[Correlation of the displacement of particles,
$C_{\vec{D}\vec{D}}(r)$, for a system under vibration.]{(Color
online) Correlation of the displacement of particles,
$C_{\vec{D}\vec{D}}(r)$, for a system under vibration. Values of
$\phi/\phi_c$ are shown in the plots. Dashed horizontal line
indicates zero correlation.} \label{Figure17}
\end{figure}

To further test the above picture, we performed the experiment with
small vertical vibrations applied to the system
(Fig.~\ref{Figure3}).  By vibrating the system, any force chains
formed before jamming are destroyed by relative slip between
particles. The vibration also helps one to demobilize the frictional
contact between particles.  If our picture of the relation between
friction and jamming transition is correct, then by adding vibration
the system should jam in one single step.  In the experiment, the
vibration is generated by a mechanical shaker in a tapping mode.
Each tap is excited by one full period of a sinusoidal wave with the
frequency $\omega=30$ Hz.  The peak-to-peak acceleration of
vibration is $\Gamma_{p-p}=1.51g$, measured by an accelerometer
attached to the cell.  Here, $g$ is the gravitational acceleration.
The amplitude of vibration $A=\Gamma_{p-p}/(2\omega^2)=0.84$ mm is
much smaller than the diameter of particles. Hence, particles only
vibrate locally around their mean position.   We tapped the cell
once every 20 s before taking an image. A time interval of 2 s is
allowed between shaking and taking an image, so the system is
stationary when the image is taken.  Six different experiments all
with low initial packing fractions are performed. As expected, none
of the experiments shows the multiple-step jamming. All the systems
show a single peak in $g_1(\phi)$.  We show a typical result in
Fig.~\ref{Figure16}. Different from the results without vibration
(Fig.~\ref{Figure8}c), the mean square displacement of particles,
$\langle D^2(\phi)\rangle$, decays monotonically
(Fig.~\ref{Figure16}).  At low packing fraction, particles have more
free room to vibrate and therefore have larger amplitude of
displacement. The displacement amplitude decreases as the packing
fraction of system increases and goes to zero when the system jams.
Furthermore, the system under vibration does not develop any cluster
structure as it approaches the jamming point. As shown in
Fig.~\ref{Figure17}, the correlation of the displacement of
particles, $C_{\vec{D}\vec{D}}(r)$, keeps roughly the same shape as
$\phi \to \phi_c$, which is clearly different from the non-vibrating
case. When the cluster structure develops, $C_{\vec{D}\vec{D}}(r)$
increases significantly (Fig.~\ref{Figure11}). This confirms the
argument that the cluster structure, and therefore the static length
scale of the jamming transition, is due to the athermal nature of
the system. Vibration thermalizes the system in a certain sense, and
therefore destroys the cluster structure.

\section{VI. Conclusions}

In this paper, we systematically investigate the jamming transition
in a 2D granular system.  We show that there is a clear structural
signature of the jamming transition at zero temperature.  The heights
of the first and second peaks of the pair correlation function,
$g_1(\phi)$ and $g_2(\phi)$, both show a maximum as the system
crosses the jamming point.  By measuring the pressure along the
boundary and the displacement of particles, we show that this
maximum coincides with the mechanical and kinematic criteria of
jamming. Therefore, the structural signature we found here is a
signature of the jamming transition.  Although the amplitude of the
peak does not diverge due to the polydispersity of our particles,
our experiment corroborates the results predicted in simulations
with ideal frictionless particles at zero temperature
\cite{Silbert2}.  This signature has already been used as a new
criterion of jamming at finite temperature, where the mechanical and
kinematic criteria of jamming are hard to measure directly
\cite{Zhang}.  Here, our measurement that shows the coincidence of
the maximum with the onset of the rigidity provides an experimental
basis for the criterion. Of particular importance is that this
structural signature exists for real granular systems with
frictional contacts.  The structural signature found in experiments
reflects the underlying singularity of the jamming transition.

Friction may cause systems to jam in a series of steps.  We found
that if the initial packing configuration is highly inhomogenous,
the system may jam in a certain direction but not in others.  This
phenomenon is directly related to the well-known force chain
structure of a granular system \cite{Jaeger}.  Our observation
provides more insight on the relation between the heterogeneous
force-chain structure and the jamming transition in the presence of
friction \cite{Cates,Alexander,O'Hern2}. We speculate that the
jamming transition of frictional system is obtained through a
continuous buckling of force chains in the different directions.

It is also useful to note that $g_1(\phi)$ is a very sensitive probe
for the jamming of system.  Even partial jamming can induce a peak
in $g_1(\phi)$.  By contrast with the kinematic criterion of
jamming, the average displacement or velocity of particles drops to
zero only at the final global jamming point.  For the mechanical
criterion of jamming, depending on where one measures the pressure
along the boundary, it may show jamming at different packing
fractions.  Therefore, the peak of $g_1(\phi)$, which can be called
the geometrical criterion, is a better jamming criterion for a
system with friction.

Due to its athermal nature, this system has a dependence on its
initial particle configuration.  It amplifies any small initial
density variation and self-organizes into clusters. A static length
scale extracted from this cluster structure reaches the system size
when the system approaches the jamming point. Hence, we show a
divergent static length scale in this nonequilibrium jamming system.

\section{Acknowledgements}

The author thanks S. Nagel and H. Jaeger for their support and
guidance. I am grateful to E. Brown, N. Keim, J. Royer, N. Xu and
L.-N. Zou for their help with the experiment and fruitful
discussions. I also thank M. van Hecke for the careful reading of
the original manuscript and many constructive comments and
suggestions on the paper. This work was supported by the NSF MRSEC
program under DMR-0820054, by the Keck Initiative for Ultrafast
Imaging at the University of Chicago and by the DOE under
DE-FG02-03ER46088.

\end{document}